# Decoherence of nuclear spins due to direct dipole-dipole interactions probed by resistively detected nuclear magnetic resonance

T. Ota[1,2], G. Yusa[1,3,4], N. Kumada[1], S. Miyashita[5], T. Fujisawa[1] and Y. Hirayama[1,2,4]

[1]*NTT Basic Research Laboratories, NTT Corporation, Atsugi, Kanagawa, 243-0198, Japan*
[2]*SORST-JST, Kawaguchi, Saitama 331-0012, Japan*
[3]*PRESTO-JST, Kawaguchi, Saitama 331-0012, Japan*
[4]*Department of Physics, Tohoku University, Sendai, Miyagi, 980-8578, Japan*
[5]*NTT-AT, Atsugi, Kanagawa, 243-0198, Japan*

We study decoherence of nuclear spins in a GaAs quantum well structure using resistively detected nuclear magnetic resonance. The transverse decoherence time $T_2$ of $^{75}$As nuclei is estimated from Rabi-type coherent oscillations as well as by using spin-echo techniques. By analyzing $T_2$ obtained by decoupling techniques, we extract the role of dipole-dipole interactions as sources of decoherence in GaAs. Under the condition that the device is tilted in an external magnetic field, we exhibit enhanced decoherence induced by the change in strength of the direct dipole-dipole interactions between first nearest-neighbor nuclei. The results agree well with simple numerical calculations.

Nuclear magnetic resonance (NMR) is attracting interest in terms of implementing quantum information processing [1]. Particularly, solid-state NMR using semiconductor materials is favored for reasons of potential scalability and possible integration with conventional electronics. However, the coherence time of nuclear spins in semiconductors is strongly influenced by various interactions between nuclei and surrounding environments. A recently developed on-chip NMR device using an integer/fractional quantum Hall system [2,3] is one of the feasible devices for studying the interactions in semiconductors. Such an all-electrical device, in which nuclear spin states in a submicron scale region are sensitively detected by the change in the longitudinal resistance, enables the coherent manipulation of nuclear spins with the help of radio frequency (rf) pulses. Moreover, the decoupling of the interactions, such as heteronuclear and electron-nuclear spin interactions, by multi-layer gate structures integrated in the device allows us to study sources of decoherence in semiconductors [4,5].

In this letter, we study decoherence of nuclear spins of $^{75}$As nuclei in a GaAs quantum well using a resistively detected NMR device that works in a fractional quantum Hall regime. We estimate transverse decoherence time $T_2^{Rabi}$ from Rabi-type coherent oscillations. $T_2^{Rabi}$ is consistent with $T_2^{echo}$ obtained by a spin echo technique, and is significantly improved by the heteronuclear and electron-nuclear spin decoupling. By analyzing decoherence rates $1/T_2^{Rabi}$, we discuss the role of the direct dipole-dipole (d-d) interactions as sources of decoherence. By performing similar analysis when the device is tilted in an external static magnetic field $B_0$, we extract enhanced decoherence due to the change in the direct d-d interaction between the first nearest-neighbor nuclei. The results are consistent with simple numerical calculations.

A cross-view of the device is schematically shown in Fig. 1(a). The device consists of a 20-nm-thick GaAs/Al$_{0.3}$Ga$_{0.7}$As quantum well structure grown on GaAs (001) substrate, and multi-layer gates are integrated in the device [3]. Back gate $V_b$ enables us to control the electron density of the two-dimensional electron gas (2DEG), which is formed in the GaAs layer. A single pair of metal split gates with a gap of 800 nm was deposited on the surface in order to squeeze the conducting 2DEG into a narrow channel. By applying a negative voltage of -0.4 V and a constant source-drain current of 10 nA, the electron-nuclear spin coupling is pronounced only in the constricted region when the electron system is set in a fractional quantum Hall state, with Landau level filling factor $v = 2/3$ [6]. An antenna gate deposited just above the split gates is used to apply rf pulses. The nuclear spin states modified by rf pulses are sensitively detected as the change in the longitudinal resistance $\Delta R_{xx}$ [3]. The $R_{xx}$ measurements were performed using standard lock-in techniques. The device is tilted by angle $\alpha$ (unit: degree) along the (1-10) axis while keeping component $B_0$ perpendicular to the 2DEG constant. The sample is mounted in a dilution refrigerator with a temperature below 70 mK.

The decoupling of the electron-nuclear spin and heteronuclear spin interactions are performed by the back and antenna gates integrated in the device, respectively [4,5]. Figure 1(b) shows resistively detected pulsed NMR spectra for $^{75}$As nuclei at $B_0$=4.27 T, $V_b$=-0.51 V and $\alpha = 0°$, obtained by sweeping the frequency of the rf pulse while measuring $\Delta R_{xx}$. The upper and lower traces correspond to the spectra with and without the electron-nuclear spin decoupling, respectively. The three peaks reveal possible single-photon transitions between neighboring spin levels $|3/2\rangle, |1/2\rangle, |-1/2\rangle$ and $|-3/2\rangle$ of $^{75}$As nuclei with nuclear spin $I$=3/2. The pulse duration is 120 μsec, which approximately corresponds to a π pulse for all transitions (We call this π pulse as a detection pulse). The detection pulse exchanges the nuclear spin population between the neighboring two levels. This leads to the change in the $z$ component of magnetization, which is measured by $\Delta R_{xx}$. For the upper spectrum, the detection pulse was irradiated after the 2DEG was completely depleted by applying a negative back gate voltage of -1.4 V. The detailed procedures are described in Refs. 4 and 5. The shift of the peak position represents the Knight shift, indicating that the Larmor frequency of $^{75}$As nuclei is modified by magnetic moments of surrounding conduction electrons. For the heteronuclear decoupling, rf continuous waves, which randomize nuclear spins of $^{69}$Ga and $^{71}$Ga nuclei, are applied to the antenna gate. Then, for $^{75}$As nuclei, the effective nuclear field produced by $^{69}$Ga and $^{71}$Ga nuclei becomes zero on average [7]. In the subsequent discussion, we use the right-hand peak *III* in Fig. 1(b) to estimate $T_2$ of $^{75}$As nuclei.

The orthodox method to evaluate $T_2$ is a spin-echo technique. Figure 2(a) shows a spin-echo peak obtained with the pulse sequence shown in the inset. Here, the durations for $\pi/2$ and $\pi$ pulses are given as 60 μsec and 120 μsec, respectively. Here, $\tau_1$ is fixed to 100 μsec. The change in $\Delta R_{xx}$ is maximum when the intervals $\tau_1$ and $\tau_2$ are equal. Figure 2(b) shows $\Delta R_{xx}$ obtained at the condition of $\tau_1 = \tau_2$ as a function of $2\tau_1$. The echo signal monotonically decreases due to decoherence as $2\tau_1$ increases. The dashed line is a single exponential decay curve fitted to the data, from which $T_2^{echo}$ can be estimated as 0.5 msec. Figure 2(c) shows Rabi-type coherent oscillations of $^{75}$As nuclei. The data are well fitted by damped oscillations indicated by the solid line. From this fitting, we estimate $T_2^{Rabi}$=0.6 msec. This value is consistent with $T_2^{echo}$, and longer $T_2^{Rabi}$ could possibly come from suppressed decoherence by the bang-bang effect [8]. In this letter, in order to study the sources of the decoherence of nuclear spins, we focus on not $T_2^{echo}$ but $T_2^{Rabi}$ because, in demonstrating quantum algorithms, lots of rf pulses have to be successively applied and $T_2^{Rabi}$ is therefore the relevant value to consider (Here, we define $T_2^{Rabi}$ as $T_2$).

Figures 3(a)-(c) show Rabi-type coherent oscillations of $^{75}$As at $B_0$=4.27 T and $\alpha = 0°$ without decoupling, with electron-nuclear spin decoupling, with electron-nuclear spin and heteronuclear decoupling, respectively. The spectrum in Fig. 3(a) is the same as that in Fig. 2(c). The dots and solid lines correspond to the experimental data and fitted curves, respectively. Under the decoupling, slower decay of the coherent oscillations is observed, indicating the suppression of decoherence. $T_2$ for the oscillations in Figs. 3(a), (b), and (c) are estimated as 0.6, 1.1, and 1.6 msec, respectively. Now we study individual sources of decoherence by comparing decoherence rate $1/T_2$ obtained by the decoupling techniques. The possible interactions between $^{75}$As nuclei and surrounding environments are direct homonuclear (*As-As*) and direct heteronuclear (*As-Ga*) d-d interactions and electron-mediated interactions (*As-e*). All of these three components are included in $1/T_2$=1.67 msec$^{-1}$ for the data in Fig. 3(a). In $1/T_2$=0.91 msec$^{-1}$ estimated from the data in Fig. 3(b), the *As-e* component is eliminated by the decoupling. The *As-e* and *As-Ga* components are excluded in $1/T_2$=0.63 msec$^{-1}$ obtained from the data in Fig. 3(c). Using these results, we estimate the contribution of *As-Ga*, *As-As*, and *As-e* to be $1/T_2$=0.28, 0.63 and 0.76 msec$^{-1}$, respectively, as summarized in the inset in Fig. 3(a). Note that the effect of possible imperfection in the decoupling process is included in the *As-As* component. The result indicates that electron-mediated interactions *As-e* are the main sources of decoherence and that the effect of *As-Ga* is smallest even though Ga nuclei are first nearest-neighbor nuclei for As nuclei in GaAs.

In order to further study decoherence induced by the direct d-d interaction, similar measurements were performed with the device tilted in $B_0$. Figures 3(d), (e), and (f) show the coherent oscillations at $B_0$=5.5 T and $\alpha = 32°$ without decoupling, with electron-nuclear spin decoupling, and with electron-nuclear spin and heteronuclear decoupling, respectively. $1/T_2$ corresponding to the oscillations in Figs. 3(d), (e), and (f) are estimated as 2.86, 1.43, and 0.67 msec$^{-1}$, respectively. $1/T_2$ without decoupling at $\alpha = 32°$ is much larger than that at $\alpha = 0°$, indicating enhanced decoherence induced by tilting the device [9]. By analyzing the data similarly, $1/T_2$ of *As-Ga*, *As-As*, and *As-e* are individually extracted as 0.67, 0.76 and 1.43 msec$^{-1}$, respectively, as summarized in the inset in Fig. 3(d). $1/T_2$ of the direct d-d interactions of *As-Ga* is largely enhanced from 0.28 to 0.67 and that of *As-As* slightly increases from 0.63 to 0.76.

The change in $1/T_2$ according to the tilt angle is explained by the Hamiltonian of the direct d-d interaction under the presence of $B_0$. The angle-dependent component in Hamiltonian of the d-d interaction between homo- and heteronuclei is represented as $\hbar^2 \gamma_{As} \gamma_{As} I_z^{As} I_z^{As} (1-3\cos^2\theta)/r^3$ and $\hbar^2 \gamma_{Ga} \gamma_{As} I_z^{Ga} I_z^{As} (1-3\cos^2\theta)/r^3$, respectively [10]. Here, $\hbar$ is Planck's constant $h$ divided by $2\pi$. $I_z^{Ga(As)}$ and $\gamma_{Ga(As)}$ are the z component of nuclear spins and the gyromagnetic ratio of $^{69}$Ga and $^{71}$Ga ($^{75}$As) nuclei, respectively. $r$ is the distance between the position of two nuclei, i.e. $^{75}$As and other nucleus, and $\theta$ is the angle between the direction of the bond and $B_0$. Figure 4 shows calculated $|\gamma_{Ga(As)} \gamma_{As} (1-3\cos^2\theta)/r^3|$ as a function of $\alpha$. Dashed lines *I* to *IV* correspond to the summation of $|\gamma_{Ga(As)} \gamma_{As} (1-3\cos^2\theta)/r^3|$ for the first to fourth nearest-neighbor nuclei, respectively. Here, $\theta$ of each nucleus at the given $\alpha$ is calculated. The intensity decreases and the curves reveal less angular dependence for farther nuclei. The curve *I+III* and *II+IV* correspond to the summed contribution of *As-Ga* and *As-As*, respectively. We compare the experimental data with the values obtained from the curves *I+III* and *II+IV*. Experimentally, $1/T_2$ of *As-Ga* changes from 0.28 to 0.76 when $\alpha$ increases from 0° to 32°. $1/T_2$ of 0.28 at $\alpha = 0°$ comes from the contribution of third nearest-neighbor nuclei since the effect of the first nearest-neighbor nuclei becomes zero. This is due to the magic angle configuration in the Zincblende crystal structure of GaAs, i.e. the angle between the bond and $B_0$ is 54.74 degrees ($\theta$ =54.74). By tilting the device, the contribution of the first nearest-neighbor nuclei is drastically enhanced. The enhanced $1/T_2$ by a factor of 2.7 is consistent with the ratio of 2.9 estimated from the calculated curve *I+III* in Fig. 4. For *As-As*, $1/T_2$ slightly changes from 0.63 to 0.67. The ratio of these $1/T_2$ is 1.06, which is again consistent with the ratio of 1.05 estimated by the curve *II+IV*. The contribution of *As-e* is largest at $\alpha = 0°$ and $\alpha = 32°$. This indicates that temporal fluctuations of conduction electrons during the coherent oscillations affect the decoherence of nuclear spins more severely than those of surrounding nuclei because the magnetic moments of electrons are larger than those of nuclei. The enhancement of $1/T_2$ of *As-e* with the device tilted might come from effects, such as enhanced electron-mediated indirect d-d interactions between As and Ga nuclei and the change in the strength of the electron-nuclear spin interactions due to the increase in Zeeman energy of electrons.

In conclusion, we studied $T_2$ of nuclear spins using the decoupling techniques that allow us to extract individual sources of decoherence. By tilting the device in $B_0$, we extracted enhanced decoherence due to the interaction between first-neighbor nuclei. These results on the d-d interactions might pave the way for the coherent control between neighboring qubits of As and Ga nuclei.

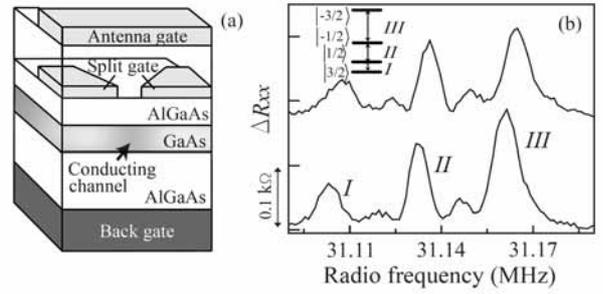

Fig.1 (a) Schematic illustration of a cross-sectional view of the device. (b) Resistively detected NMR spectrum of $^{75}$As nuclei at 4.27 T. The upper and lower traces correspond to the spectra with and without the electron-nuclear spin decoupling, respectively. Inset: Possible schematic energy level diagram of nuclear spins for *I*=3/2. The transitions *I*, *II*, and *III* in the inset correspond to the peaks labeled as *I*, *II*, and *III*.

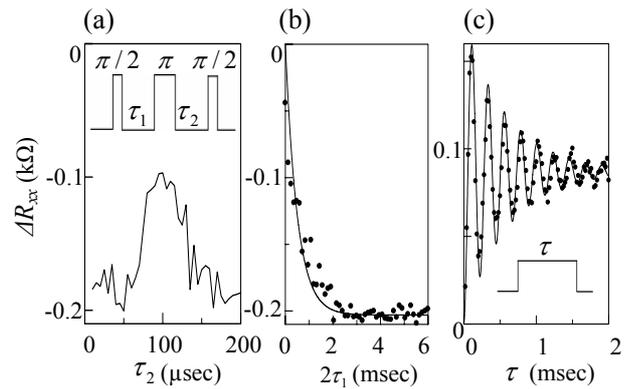

Fig.2 (a) Spin-echo peak as a function of $\tau_2$. $\tau_1$ is 100 μsec. Inset: Illustration of the pulse sequence for spin-echo experiment. (b) Spin-echo spectrum as a function of $2\tau_1$. (c) Rabi-type coherent oscillations. Inset: Illustration of the pulse sequence for the coherent oscillations. The solid lines in Figs. 2(b) and (c) are fitted curves.

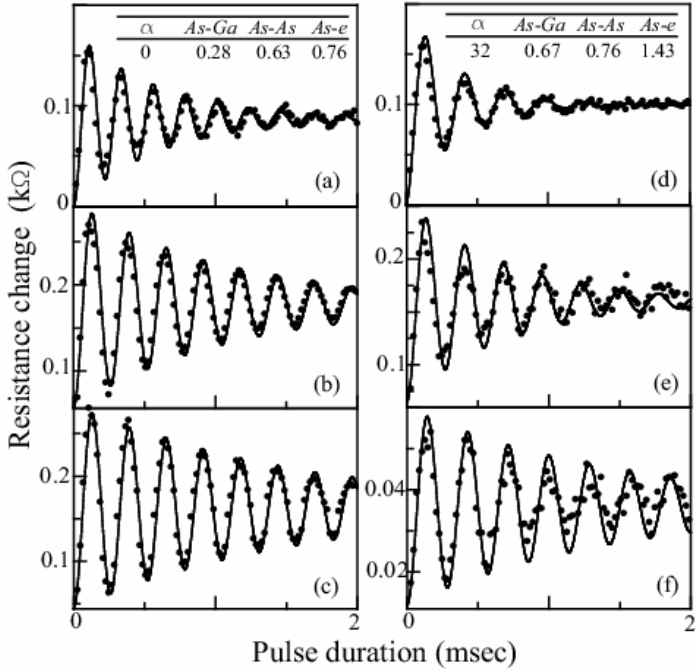

Fig.3 (a)-(c) Rabi-type coherent oscillations of $^{75}$As nuclei at $B_0$=4.27 T and $\alpha = 0°$ without decoupling, with electron-nuclear spin decoupling, and with electron-nuclear spin and heteronuclear decoupling, respectively. (d)-(f) Similar oscillations at $B_0$=5.5 T and $\alpha = 32°$ without decoupling, with electron-nuclear spin decoupling, and with electron-nuclear spin and heteronuclear decoupling, respectively. The solid lines are fitted curves.

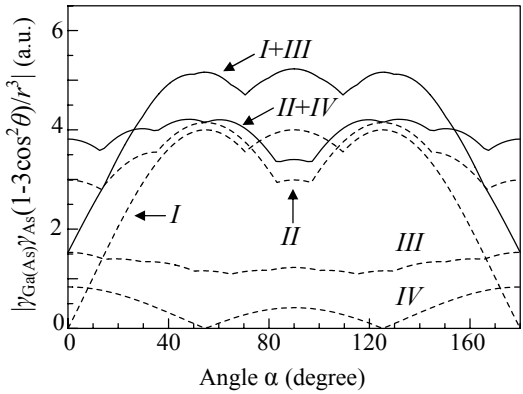

Fig.4 Calculated $|\gamma_{Ga(As)}\gamma_{As}(1-3\cos^2\theta)/r^3|$ normalized by $\gamma_{As}\gamma_{As}$ as a function of $\alpha$. Here, $r$ is a unit of $\sqrt{3}/4a$, which is the distance between the first nearest-neighbor nuclei ($a$ is a lattice constant of GaAs). Dashed lines $I$ to $IV$ indicate $|\gamma_{Ga(As)}\gamma_{As}(1-3\cos^2\theta)/r^3|$ corresponding to the first to fourth nearest-neighbor nuclei, respectively. The curve $I+III$ and $II+IV$ represent the summed contribution of As-Ga and As-As, respectively.